\documentstyle{aipproc2}

\input epsf

\begin{document}
\title{Structure formation with strings plus 
inflation: a new paradigm}

\author{Jo\~ao Magueijo$^1$, 
Carlo Contaldi$^{1}$, and Mark Hindmarsh$^2$}
\address{$^1$Theoretical Physics, The Blackett Laboratory, 
Imperial College, Prince Consort Road, London, SW7 2BZ, U.K.\\ 
$^2$ Centre for Theoretical Physics, University of Sussex,  
Brighton BN1 9QJ, U.K.} 


\maketitle

\begin{abstract} 
Recent developments in inflation model building, based on supersymmetry,
have produced compelling models in which strings are 
produced at the end of inflation.
In such models the cosmological perturbations are 
seeded both by the defects and by the quantum fluctuations. 
We show that such models produce 
qualitatively new and desirable predictions for CMB anisotropies and
the CDM power spectrum. This remark should put an end to the long
term animosity between defect and inflationary scenarios of structure
formation. 
\end{abstract}

\section{Introduction}

Structure formation theories fall broadly within two classes:
inflation \cite{scdm} and defects \cite{csreviews}. In inflationary scenarios
the structure of the Universe originates from microphysical
quantum fluctuations, which get stretched to cosmological scales
by inflationary expansion. In topological defect scenarios 
as the Universe cools down, high temperature
symmetries are spontaneously broken. Remnants of the unbroken phase, 
called topological defects, may survive the transition, and later 
seed fluctuations in the CMB and CDM. 

A major drawback of inflationary theories 
is that they are far-removed from particle physics
models. Attempts to improve on this state of affairs have been made
recently, resorting to supersymmetry 
\cite{Cas+89,DTermInfl,rachel,LytRio98,Cop+94,Tka+98,sug}.
In these models one identifies
flat directions in the potentials, which are enforced by a (super)symmetry. 
Such flat directions produce ``slow-roll inflation''. 
In order to stop inflation one must tilt the potential, allowing 
for the fields to roll down.  
In so-called D-term supersymmetric inflationary scenarios, 
inflation stops with a symmetry-breaking phase transition, 
at which a U(1) symmetry is spontaneously broken, leading to the 
formation of cosmic strings.  This is only the most 
natural of a whole class of models of so-called hybrid inflation.
Hence a network of cosmic strings is formed at the end of inflation.

If one takes the standard cosmology (sCDM), 
of $\Omega=1$, $\Omega_\Lambda=0$,   
$\Omega_b = 0.05$ and $H_0 = 50$ km s$^{-1}$ Mpc$^{-1}$, one finds
that neither strings or inflation fit the COBE normalized large
scale structure power spectrum. 
However, the failings of the inflationary and defect sCDM models are  
to a certain extent complementary, and an obvious question is whether 
they can help each other to improve the fit to the data.

Using our recent calculations for local cosmic strings \cite{chm1} and the by 
now familiar inflationary calculations \cite{cmbfast}, we are able to 
demonstrate that the answer is yes \cite{chm2}. Even with Harrison-Zeldovich
initial conditions and no inflation produced gravitational waves,
the large-angle CMB spectrum is mildly tilted, as preferred by COBE
data \cite{kris}. The CMB spectrum then rises into a thick Doppler bump, 
covering the region $\ell=200-600$, modulated by soft secondary 
undulations. More importantly the standard CDM anti-biasing problem is cured, 
giving place to a slightly biased scenario of galaxy formation. The
cosmic string biasing problem is also cured. 

Similar results have been reported by two other groups \cite{stinf1,stinf2}.

\section{Model building}
The general features of structure formation with strings plus inflation
do not depend on the concrete underlying inflationary model.
We illustrate these models by considering
the $D$-term inflation model in which the strings plus inflation
scenario finds an attractive expression. 

To begin with, we 
define the reduced Planck mass $M = 1/\sqrt{8\pi G}$.  We 
recall that a supergravity theory is defined by two functions of the  
chiral superfields $\Phi_i$: the function $G(\bar\Phi,\Phi)$, 
which is related to the K\"ahler potential $K(\bar\Phi,\Phi)$ 
and the superpotential $W(\Phi)$ 
by $G=K+M^2\ln|W|^2/M^6$, and the 
gauge kinetic function $f_{AB}(\bar\Phi,\Phi)$.  
The scalar potential $V$ is composed of two terms, 
the $F$-term 
\begin{equation} 
V_F = M^2 e^{G/M^2}\left(G_i (G^{-1})^i_{\,j} G^j - 3 {M^2} \right)  
\end{equation}
and the $D$-term
\begin{equation}
V_D = \frac{1}{2} g^2 {\mathrm{Re}} f_{AB}^{-1} D^A D^B 
\end{equation} 
where  $g$ is the U(1) gauge  
coupling, $G^i = \partial G/\partial \Phi_i$, and 
$G_i = \partial G/\partial \bar\Phi^i$.  The function   
$D^A$ i given by 
\begin{eqnarray} 
D^A &=& G^i(T^A)_i^{\,j}\phi_j + \xi^A, 
\end{eqnarray} 
where the Fayet-Iliopoulos terms $\xi^A$, which we take to be 
positive, can be non-zero only for those $(T^A)_i^{\,j}$ which are U(1) 
generators.  We see that in order to have a positive potential energy density,  
either the $F$ term or the $D$ term must be non-zero.  In order to have
inflation, there must be a region in field space where the slow-roll
conditions $\epsilon \equiv \frac12 M^2 |V^i/V|^2 \ll 1$ and 
$|\eta| \equiv |\min {\mathrm eig} \, M^2 V^i_{\,j}/V| \ll 1$ 
are satisfied, where by the notation in the 
second condition we mean that the smallest 
eigenvalue of the matrix is much less than unity.
In $D$-term inflation, the conditions are satisfied because 
the fields move along a trajectory for which  
$\exp(G/M^2)$, $G^i$ and $G^i(T^A)_i^{\,j}\phi_j$ all vanish, leaving a tree-level  
potential energy density of $g^2\xi^A\xi^A/2$.  Thus the potential is  
completely flat before radiative corrections are taken into account.  At the  
end of inflation, if the fields are to relax to the supersymmetric minimum with  
$D^A+\xi^A = 0$, the U(1) gauge symmetries are necessarily broken,
assuming their  
corresponding Fayet-Iliopoulos terms are non-zero.  Thus strings are  
inevitable: the only question is how much inflation there is before the  
fields attain the minimum.

\section{Calculations}
The spectrum of the perturbations from $D$-term inflation is  
calculable \cite{rachel}, and can be expressed in terms of $N$, 
the number of $e$-foldings between the horizon exit of cosmological  
scales today and the end of inflation, which occurs at $|\eta| = 1$. 
One finds  
\begin{equation} 
{\ell(\ell+1)C_\ell^{\mathrm{I}}\over 2\pi T_{\mathrm{CMB}}^2 }\simeq 
\frac14 |\delta_H(k)|^2 \simeq  
\frac{(2N+1)}{75} \left(\frac{\xi^2}{M^4}\right), 
\end{equation} 
where $ T_{\mathrm{CMB}}= 2.728 K$ 
is the temperature of the microwave  
background, and $\delta_H(k)$ is the matter perturbation amplitude at 
horizon crossing. The corrections to this formula, which is zeroth order in  
slow roll parameters, are not more than a few per cent.   
The inflationary fluctuations in this model are almost 
scale-invariant (Harrison-Zeldovich) and have a negligible tensor
component \cite{DTermInfl}. 

The string contribution is uncorrelated with the inflationary  
one, and is proportional to $(G\mu)^2$, where $\mu$ is 
the string mass per unit length, given by $\mu = 2\pi\xi$.  We can write  
it as  
\begin{equation} 
{\ell(\ell+1)C_\ell^{\mathrm{S}}\over 2\pi T_{\mathrm{CMB}}^2} =  
 \frac{{\cal A}^{\mathrm{S}}(\ell)}{16}\left(\frac{\xi^2}{M^4}\right), 
\end{equation} 
where the function ${\cal A}^{\mathrm{S}}(\ell)$ gives the amplitude of the  
fractional temperature fluctuations in units of $(G\mu)^2$.  Allen et al.\  
\cite{all+} report ${\cal A}^{\mathrm{S}}(\ell) \simeq 60$ on large  
angular scales, with little dependence on $\ell$.  Our simulations give  
${\cal A}^{\mathrm{S}}(\ell) \simeq 120$, with a fairly strong tilt.
The source of the difference is not altogether clear: our simulations 
are based on a flat  
space code which neglects the energy losses of the strings through  
Hubble damping.  The simulations of Allen et al.\ do include Hubble damping, 
which would tend to reduce the string density and hence the normalisation. 
However, they have a problem of lack of dynamic range, 
and therefore may be missing  
some power from strings at early times, and therefore higher $\ell$.
 
Jeannerot \cite{rachel} took the Allen--Shellard normalisation 
and $N\simeq 60$, and found that the  
proportion of strings to inflation is roughly $3:1$.  
With our normalisation, the approximate ratio is $4:1$.
In any case this ratio is far  from a robust prediction in strings 
plus inflation
models, as it depends on the number of $e$-foldings,
and the string normalisation, both of which are uncertain.  
We will therefore leave it as a free parameter. For definiteness
we shall parametrize the contribution due to strings and inflation
by the strings to inflation ratio $R_{\mathrm{SI}}$, 
defined as the ratio in $C_\ell$ at $\ell=5$, that is
$R_{\mathrm{SI}}=C_5^S/C_5^I$. 

It is curious to note that the number of e-foldings required for
solving the flatness problem still leaves room for tuning
$R_{\mathrm{SI}}$ between nearly 0 and 1. 

\section{Results}

In Figs. 1 and 2 we present power spectra in CMB and CDM produced 
by a sCDM scenario, by cosmic strings, and by strings plus inflation. 
We have assumed the traditional choice of parameters, setting the Hubble 
parameter
$H_0=50$ km sec$^{-1}$ Mpc$^{-1}$, the baryon fraction to
$\Omega_b=0.05$, and  assumed a flat geometry, no cosmological
constant, 3 massless neutrinos, standard recombination,
and cold dark matter. The inflationary perturbations have a 
Harrison-Zeldovich or scale invariant spectrum, and the amount of
gravitational radiation (tensor modes)
produced during inflation is assumed to be negligible.

\begin{figure}
\centering
\leavevmode\epsfysize=7cm \epsfbox{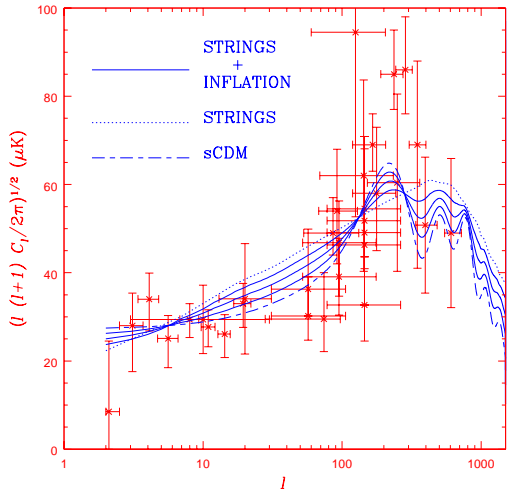}\\ 
\caption[flatness]{\label{fig1}The CMB power spectra predicted by cosmic strings, sCDM, 
and by inflation and strings with
$ R_{\mathrm{SI}}=0.25,0.5,0.75.$ 
The large angle
spectrum is always slightly tilted. The Doppler peak becomes a thick
Doppler bump at $\ell=200-600$, modulated by mild undulations.} 
\end{figure} 

\begin{figure}
\centering
\leavevmode\epsfysize=7cm \epsfbox{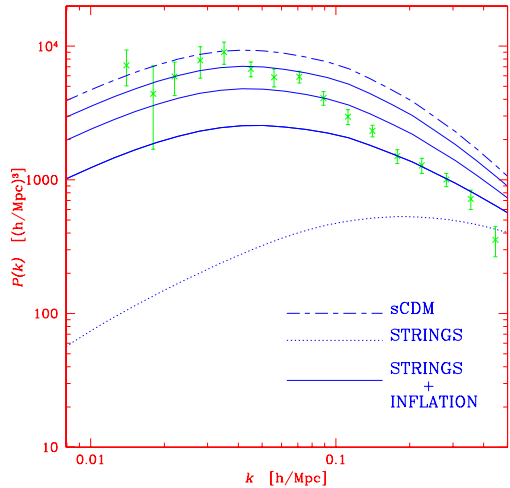}\\ 
\caption[flatness]{\label{fig2} The  power spectra in CDM fluctuations 
predicted by cosmic strings, sCDM, 
and by inflation and strings with $R_{\mathrm{SI}}=0.25,0.5,0.75.$
We have also superposed the power spectrum as inferred from surveys
by Peacock and Dodds \cite{pdodds}.} 
\end{figure} 

We now summarise the results.
\begin{itemize}
\item 
The CMB power spectrum shape in these models
is highly exotic. The inflationary 
contribution is close to being Harrison-Zeldovich. Hence it produces
a flat small $\ell$ CMB spectrum. The admixture of strings, 
however, imparts a tilt.
Depending on $R_{\mathrm{SI}}$ one may tune the CMB plateau tilt between 1
and about 1.4, without invoking primordial tilt and inflation
produced gravity waves.

\item
The proverbial inflationary Doppler peaks are transfigured in these
scenarios into a thick Doppler bump, covering the region  
$\ell=200-600$. The height of the peak is similar for sCDM and strings,
with standard cosmological parameters. 
The Doppler bump is modulated by small undulations, 
which cannot truly be called secondary peaks. 
By tuning $R_{\mathrm{SI}}$ one may achieve any degree
of secondary oscillation softening. This provides a major
loophole in the argument linking inflation with secondary
oscillations in the CMB power spectrum \cite{andbar,inc}. 
If these oscillations were not
observed, inflation could still survive, in the form of the models
discussed in this Letter.

\item
In these scenarios the LSS of 
the Universe is almost all produced by inflationary fluctuations. 
However COBE scale CMB anisotropies are due to both strings and  
inflation. Therefore COBE normalized CDM fluctuations are  
reduced by a factor $(1+R_{\mathrm{SI}})$ in strings plus inflation scenarios. 
This is equivalent to multiplying 
the sCDM bias by ${\sqrt{1+R_{\mathrm{SI}}}}$ on all scales, except
the smallest, where the string contribution may be
non negligible. Given that 
sCDM scenarios produce too much structure on small scales
(too many clusters)  
this is a desirable feature. 
\end{itemize}

\section{Praise for the model}
``Strings plus inflation'' are interesting first of all as an inflationary
model. Its ``flat potential'' is not the result of a finely tuned coupling
constant, but the result of a symmetry. Hence in some sense
these models achieve inflation without fine tuning. The only free parameters
are the number of inflationary e-foldings, and the scale of symmetry
breaking. These parameters also fix the absolute (and therefore relative)
normalizations of string and inflationary fluctuations.

``Strings plus inflation''  models are also pervaded by a higher component 
of particle physics, when compared with other
inflationary models. 

The structure formation paradigm resulting from this scenario is highly
exotic and worth considering just by itself. Regarded in abstract, 
structure formation may be due to two types of mechanism: 
active and passive perturbations. Passive fluctuations are due 
to an apparently acausal imprint in the initial conditions of the standard 
cosmic ingredients, which are then left to evolve by themselves. 
Active perturbations are due to an extra cosmic component, which 
evolves causally (and often non-linearly), and drives perturbations 
in the standard cosmic ingredients at all times.  
Inflationary fluctuations are passive. Defects are the  
quintessential active fluctuation. 
A scenario combining active and passive perturbation would 
bypass  most of the current wisdom on what to expect in either 
scenario. It is believed that the presence or absence of secondary 
Doppler peaks in the CMB power spectrum tests the very fundamental nature
of inflation, whatever its guise \cite{andbar}.
In the mixed scenarios we shall consider inflationary scenarios
could produce spectra with any degree of secondary oscillation
softening.

The combination of these two scenarios smoothes 
the hard edges of either separate component, leaving 
a much better fit to LSS and CMB power spectra. We illustrated this
point in this review, but left
out a couple of issues currently under investigation
which we now summarise.

The CDM power spectrum in these scenarios has a break at very small
scales, when string produced CDM fluctuations become dominant over
inflationary ones. This aspect was particularly emphasized in
\cite{stinf2}, and there is some observational evidence in favour
of such a break. An immediate implication of this result is that
it is easier to form structure at high redshifts \cite{steidel,lalfa}. 
In \cite{bmw} it is shown that even with Hot Dark Matter, these
scenarios produce enough damped
Lyman-$\alpha$ systems, to account for the recent high-redshift
observations.

Another issue currently under investigation is the timing of structure
formation \cite{steidel}. Active models drive fluctuations
at all times, and therefore produce a time-dependence
in $P(k)$ different from passive models. The effect is subtle,
but works so as to slow down structure formation. Hence for
the same normalization nowadays there is more structure at high
redshifts in string scenarios. 

Overall we end up with a picture in which the CMB is produced 
by both strings and inflation, the current large scale structure
of the Universe is  produced by inflation except on the very small 
scales, but most of the structure at high redshift is produced by strings.

In such models there
would also be intrinsic non-Gaussianity at the scale of clusters,
with interesting connections with the work of \cite{james}.

\section*{Acknowledgments}

JM and CC thank the organizers for an excellent meeting. 
We acknowledge financial support from the Beit Foundation (CC),  
PPARC (MH), and the Royal Society (JM), and also from the EC
(JM and CC).


\begin{references}
\bibitem{scdm} P. Steinhardt, {\it Cosmology at the
Crossroads}, {\it Proceedings of  the Snowmass Workshop on
Particle Astrophysics and Cosmology}, E. Kolb and R.Peccei,
eds. (1995) 
\bibitem{csreviews} A. Vilenkin and E.P.S. Shellard, Cosmic Strings and Other  
Topological Defects (Cambridge Univ. Press, Cambridge, 1994); 
M. Hindmarsh and T.W.B. Kibble, Rep. Prog. Phys {\bf 55}, 478 (1995). 
\bibitem{Cas+89} J.A. Casas, J.M. Moreno, C. Mu\~noz, and M. Quir\'os, 
{ Nucl. Phys.} {\bf B328} 272 (1989).
\bibitem{DTermInfl} E. Halyo, { Phys. Lett.} {\bf B387} 43 (1996); 
P. Bin\'etruy and G. Dvali, { Phys. Lett.} {\bf B388} 241 (1996). 
\bibitem{rachel} R. Jeannerot, { Phys. Rev.} {\bf D56} 6205 (1997). 
\bibitem{LytRio98} D.H. Lyth and A. Riotto, hep-ph/9807278.
\bibitem{Cop+94} E.J. Copeland et al
{ Phys. Rev.} {\bf D49} 6410 (1994).
\bibitem{Tka+98} I. Tkachev et al,
hep-ph/9805209.
\bibitem{sug}T. Kanazawa et al, hep-ph/9803293.
\bibitem{chm1} C. Contaldi, M. Hindmarsh, J. Magueijo, 
{\it Phys. Rev. Lett}, {\bf 82} (1999).
\bibitem{cmbfast}U. Seljak and M. Zaldarriaga, {  Astrop. J.} 
{\bf 469}, 437 (1997). 
\bibitem{chm2} C. Contaldi, M. Hindmarsh, J. Magueijo, 
{\it Phys. Rev. Lett}, IN PRESS, astro-ph/9809053.
\bibitem{kris}K.Gorski, Proceedings of the XXXIst Recontres
     de Moriond, 'Microwave Background Anisotropies'
\bibitem{stinf1}P. Avelino, R. Caldwell, C. Martins,
astro-ph/9809130.
\bibitem{stinf2} R. Battye, J. Weller, astro-ph/9810203.
\bibitem{all+}B. Allen et al, { Phys. Rev. Lett.} {\bf 77} 3061 (1997). 
\bibitem{andbar}J.Barrow and A.Liddle, { Gen. Rel. Grav.}
 {\bf 29} 1503 (1997).
\bibitem{inc}A. Albrecht et al, {  Phys. Rev. Lett} {\bf 76} 
1413 (1996); 
J.Magueijo et al, { Phys. Rev. Lett} {\bf 76} 2617 (1996). 
\bibitem{pdodds}J. Peacock and S. Dodds,  
{  M.N.R.A.S.} {\bf 267} 1020 (1994).
\bibitem{steidel}C. Steidel et al, astro-ph/9708125.
\bibitem{lalfa} L.J. Storrie-Lombardi, R.G. McMahon, and M.J. Irwin, 
M.N.R.A.S. {\bf 283} L79 (1996).
\bibitem{bmw}R. Battye, J. Magueijo, and J. Weller, in preparation. 


\bibitem{james}J. Robinson, E. Gawiser, J. Silk, astro-ph/9805181.



\end{references}
\end{document}